\documentclass[twocolumn,aps,pre,showpacs]{revtex4}
\usepackage[latin1]{inputenc}
\usepackage{graphicx}
\usepackage{latexsym}
\usepackage{amssymb}
\usepackage{amsmath}

\begin{document}
\def\d{{\rm d}}
\def\ex{{\rm e}}
\def\im{{\rm i}}
\def\U{{\bf U}}
\def\x{{\bf x}}
\def\A{{\bf A}}
\def\W{{\bf W}}
\def\k{{\bf k}}
\def\R{{\bf R}}
\def\p{{\bf p}}
\def\E{{\bf E}}
\def\pinf{{p_{\scriptscriptstyle\infty}}}
\def\smalze{{\scriptscriptstyle (0)}}
\def\smalun{{\scriptscriptstyle (1)}}
\def\smalp1{{\scriptscriptstyle (p-1)}}
\def\smalR{{\scriptscriptstyle{\rm R}}}
\def\smalN{{\scriptscriptstyle{\rm N}}}
\def\zmin{{z_{\scriptscriptstyle\min}}}
\def\zmax{{z_{\scriptscriptstyle\max}}}
\def\sec{{\rm sec}}
\def\smalrm1{{\scriptscriptstyle (r-1)}}
\def\bOmega{{\boldsymbol{\Omega}}}
\def\bsigma{{\boldsymbol{\sigma}}}
\def\rhoS{\rho_{\scriptscriptstyle S}}
\def\rhoR{\rho_{\scriptscriptstyle R}}
\def\beq{\begin{equation}}
\def\eeq{\end{equation}}
\font\brm=cmr10 at 24truept
\font\bfm=cmbx10 at 15truept

\title{Return times for stochastic processes with power-law scaling}
\author{Piero Olla}
\affiliation{ISAC-CNR and INFN, Sez. Cagliari, I--09042 Monserrato, Italy.}
\date{\today}

\begin{abstract}
An analytical study of the return time distribution of extreme events
for stochastic processes with power-law correlation has been carried on.
The calculation is  based on 
an $\epsilon$-expansion in the correlation exponent:
$C(t)=|t|^{-1+\epsilon}$. The fixed point of the theory is 
associated with stretched exponential scaling of the 
distribution; analytical expressions, valid in the 
pre-asymptotic regime, have been provided. 
Also the permanence time distribution appears to be characterized 
by stretched exponential
scaling.
The conditions for application of the theory to non-Gaussian processes
have been analyzed and the relations with the issue of return times in 
the case of multifractal measures have been discussed.

\end{abstract}

\pacs{02.50.Ey,05.40.Ca,05.45.Tp,89.75.-k} 
\maketitle

\section{INTRODUCTION}
Calculating the return time statistics of rare events in stochastic processes, 
is one of the classical problems in probability theory. The applications are
widespread, one of the most direct being the determination of safety margins against 
catastrophic events like floods and earthquakes. In other fields, like statistical 
mechanics and information theory, the statistics of return times plays an 
important role, being intimately connected with the way a system looses memory 
of its initial conditions.

Starting from the work of D\"oeblin \cite{doeblin40}, and that of Bellman and Harris 
on Markov chains \cite{bellmann51}, one of the known results is that, if the
system correlations decay sufficiently fast, the distribution of the return 
times of asymptotically rare events will tend to be exponential (see \cite{abadi01} 
for recent references).  

More recently, motivated by the observation that a wide variety of experimental 
records present long-time correlations (see \cite{bunde03} and references therein)
there has been growing interest in the case the stochastic process is power-law correlated,
with exponent small enough for the correlation time to be infinite.
An interesting result is that, in this case, 
the return time distribution of extreme events appears to be well fitted by a 
stretched exponential, with exponent equal to the power in the correlation decay, 
rather than by a simple exponential \cite{bunde03,altmann05,eichner06,pennetta06}.

Actually, seen in the light of the classical work
by Newell and Rosenblatt \cite{newell62} on the probability of no zero crossing
for power-law correlated Gaussian processes, this result is not really surprising.
If the rare event is associated with the crossing of a high threshold by 
a stochastic variable, the return time probability will be the no-crossing probability, 
for an initial condition in which the variable is right below threshold.

The central idea in \cite{newell62} is that threshold crossings will be less likely
for processes with longer correlations. 
Comparing with processes with known return time distribution (typically, a superposition 
of an Ornstein-Uhlembeck process and a stochastic variable)
Newell and Roseblatt were able to prove that the 
no zero-crossing probability for $y$ is bounded from above and from below by stretched 
exponentials.

In principle, the approach in \cite{newell62} could be extended to the case of
a threshold different from zero, allowing to conclude that stretched exponential
is the correct asymptotic scaling of the return probability for large values of its
argument. This leaves open, however, some important questions.
It would be of obvious interest to tighten the inequalities in \cite{newell62},
fixing the values of the prefactors in the stretched exponential scaling.
It would also be interesting to have some idea of how and 
when 
(and perhaps why)
the asymptotic regime is reached, and if Gaussianity is really an
essential hypothesis.

Purpose of this paper is to present an
analytical treatment of these isssues, based on
renormalized perturbation theory and an
$\epsilon$-expansion in the correlation exponent:
$\langle y(t)y(0)\rangle\sim |t|^{-1+\epsilon}$;
this is basically an expansion around the transition 
to infinite correlation time. The expansion
will turn out to work well also for rather large
values of $\epsilon$, providing
in the Gaussian case, valid approximate 
expressions for the return time distributions.

Among the other things, the analysis will point out the dominance of 
transient behaviors, in any range of practical interest for the return times.
It will also point out that, perhaps contrary to intuition, 
the return time distribution is less sensitive to the extreme statistics
of the process than to its correlation structure, in particular that of the
correlations between scales. This will allow extension of the results
to rather generic non-Gaussian processes.

This paper is organized as follows. In Section II, the main definitions and results
are recalled. In Section III, it is shown
how the long memory of the process is associated with secular behaviors in the evolution
equations for the exit probabilities. Section IV contains the main results of the
paper, and analytic expressions for the return time distribution are provided in the
form of a renormalized $\epsilon$-expansion. Sections V and VI focus again on 
the relation between the statistics of the stochastic process and that of the
return times, extending the results to the case of non-Gaussian processes.
Section VII is devoted to the statistics of permanence 
above threshold. Section VIII contains the conclusions.

\section{RETURN, PERMANENCE AND EXIT}
\label{section2}
Let us consider a
unit variance and zero mean, stationary Gaussian process $y(t)$, 
with correlation $C(t)=\langle y(t)y(0)\rangle$ decaying like a power-law
at sufficiently long time separations:
\beq
C(t)=\tilde A\, |t|^{-1+\epsilon},
\quad
|t|>\tau_0,
\quad
0\le\epsilon< 1.
\label{eq1}
\eeq
This corresponds to the energy spectrum:
\beq
C_\omega=\int\d t\ex^{\im\omega t}C(t)\sim \omega^{-\epsilon},
\label{eq2}
\eeq
which is a common occurrence in several physical systems \cite{noise}.
Notice also that the scaling in Eq. (\ref{eq1}) is that of the velocity of a 
superdiffusive particle: $y=\dot r$, $\langle |r(t)-r(0)|^2\rangle\sim |t|^{2-\epsilon}$.
For $\epsilon<0$, in turn, the correlation time would be $O(\tau_0)$, and 
$\langle |r(t)-r(0)|^2\rangle\sim |t|$, like in the case of a Brownian particle. For 
$\epsilon>1$, Eq. (\ref{eq2}) would give the spectrum of a signal 
with local anomalous diffusive behavior
$\langle |y(t)-y(0)|^2\rangle\sim |t|^{-1+\epsilon}$.

\begin{figure}
\begin{center}
\includegraphics[draft=false,width=6.5cm]{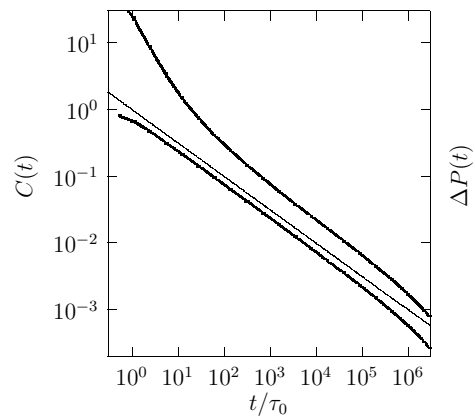}
\caption{
Scaling of the correlation function $C(t)$ (lower line) and the
probability difference $\Delta P(t)=P(y(t)>q|y(0)>q)-P(y>q)$ (upper line),
for $\epsilon=0.5$ and $N=24$. The thin line is $t^{-0.5}$.
The correlation function is obtained from Eq. (\ref{eq3}) using
$C(t)=\sum_n\langle x_n(t)x_n(0)\rangle$; $\Delta P(t)$ is obtained
using this expression for $C(t)$ and Eqs. (\ref{eq10}-\ref{eq11}).
}
\label{extrfig1}
\end{center}
\end{figure}

The Gaussian process $y(t)$ could be generated numerically, e.g. by the algorithm  
described in \cite{biferale97}, approximating $y(t)$ by
a superposition of independent Ornstein-Uhlenbeck processes $x_n(t)$, 
$n=0,1,...N$. A correlation like the one in Eq. (\ref{eq1}) could be
obtained setting for the correlation time and variance of $x_n$:
\beq
\tau_n=\tau_0 2^n
\quad {\rm and}\quad
\sigma_{x_n}^2=\frac{(1-2^{\epsilon-1})2^{(\epsilon-1)n}}{1-2^{(N+1)(\epsilon-1)}}.
\label{eq3}
\eeq
The lower line in Fig. \ref{extrfig1} is the correlation profile obtained with
this technique, for $\epsilon=0.5$ and $N=24$.

We identify an extreme event by the condition $y>q$, where $q$ is a sufficiently
large threshold for the process and introduce two occurrence times:
the permanence time $S_q$ of the variable $y$ above threshold, and 
its counterpart below threshold, the 
return time $R_q$ to the event, which coincides with the first exit 
time from $y<q$, for initial condition $y(0)=q$.
The averages $\bar R_q$ and $\bar S_q$, can be related to the event 
probability $P(y>q)$ by means of the general relation (Kac theorem \cite{kac47}):
\beq 
\bar S_q/\bar R_q\simeq P(y>q).
\label{eq3.1}
\eeq
We are interested in the occurrence time distributions
$P(S_q>t)$ and $P(R_q>t)$. We focus first on the return time distribution 
$P(R_q>t)$ and we can write:
\beq
P(R_q>t)=P(y(\tau)<q,\tau\in [0,t]|y(0)=q),
\label{eq4}
\eeq
which coincides with the no-exit probability for initial condition $y(0)=q$.
This probability is the integral from $\tau=0$ to $\tau=t$ of the probability
current across $q$. If $y(t)$ obeyed a stochastic differential equation, the
whole problem would reduce to solution of a Fokker-Planck equation with absorbing
boundaries at $q$ \cite{schuss}. 

In general, calculation of the current requires knowledge
of the profile near $q$ of the conditional PDF (probability density function)
$\rho(y(t)|y(\tau)<q,\tau\in [0,t];y(0)=q)$. Exponential scaling of $P(R_q>\tau)$
would be associated with a current $-\gamma P(R_q>t)$, with $\gamma$ a constant,
determined by the limit form of
$\rho(y(t)|y(\tau)<q,\tau\in [-\infty,t])$, when the initial condition
time is sent to $-\infty$. In the case of a long correlated stochastic
process, it will appear that this limit form is associated with zero 
current and that the approach to the limit is what generates
the anomalous scaling of the return time distribution.

\section{THE EFFECT OF MEMORY}
The problem becomes more tractable if sampling is carried on at discrete times 
$t_k=k\Delta$; indicate then $y_k=y(t_k)$ and define $A=\tilde A\Delta^{-1+\epsilon}$,
so that $C(t_k)=A|k|^{-1+\epsilon}$.
Notice that for $\Delta\gtrsim\tau_0$, the statistics would become itself $\Delta$ dependent,
and, for large $\Delta$: $\bar S_q(\Delta)\sim\Delta$; from Kac theorem
we would then have: $\bar R_q(\Delta) \sim\Delta/P(y>q)$.
Normalizing times with respect to $\Delta$, the discrete version of Eq. (\ref{eq4})
will read:
$$
P(R_q>n)=P(y_k<q,k=1,...n|y_0>q),
$$
which can be expressed in terms of the conditional probabilities:
\beq
P(y_k<q,k=1,...n|y_0>q)=\prod_{k=1}^n[1-P_k(k)],
\label{eq5}
\eeq
with $P_k(k)=P(y_k>q|y_l<q,l=1,..k-1;y_0>q)$ the return probability  
at time $t_k$ conditioned to being a first return. More in general, introduce the 
return probability conditioned to no exits before a time $t_l\le t_n$:
\beq
P_l(n)=P(y_n>q|y_j<q,j=1,...l-1;y_0>q),
\label{eq6}
\eeq
and the return probability
conditioned to exits at times $t_{k_1}<t_{k_2}<...t_{k_p}$, and no exits before a time
$t_l\le t_{k_1}$:
\beq
P_l(n|\k)\equiv P_l(n|y_{k_i}>q,i=1,...p).
\label{eq7}
\eeq
From here, an exact recursion relation can be derived, describing the evolution of the
return probability from the initial condition at $n=0$.
Equations (\ref{eq6}-\ref{eq7}) provide us, in fact, with the following identities:
$$
\begin{array}{ll}
P_{n+1}(m)&=P_n(m|y_n<q)
\\
&=[1-P_n(n)]^{-1}P_n(y_m>q,y_n<q)
\\
&=[1-P_n(n)]^{-1}[P_n(m)-P_n(y_{n,m}>q)]
\\
&=[1-P_n(n)]^{-1}[P_n(m)-P_n(n)P_n(m|n)],
\end{array}
$$
which can be rearranged to give:
$$
P_{n+1}(m)=P_n(m)
+\hat P_n(n)
[P_n(m)-P_n(m|n)],
$$
where $\hat P_n(n)=P_n(n)[1-P_n(n)]^{-1}$.  In a certain sense,
this is the analog, for a discrete non-Markovian process, of the Fokker-Planck 
equation with absorbing boundary conditions discussed previously,
and can be iterated to give:
\beq
P_n(m)=P_1(m)+\sum_{l=1}^{n-1}\hat P_l(l)[P_l(m)-P_l(m|l)].
\label{eq8}
\eeq
The physical meaning of this equation is to quantify how many of the phase points
above threshold at times $l=1,...n-1$ should be subtracted from the probability mass, that,
without taking into account the condition of first return, would be above threshold 
at time $m$.

In order to solve Eq. (\ref{eq8}), we need to know the function $P_l(m|l)$,
which is essentially the second return probability. Repeating the same
steps leading to Eq. (\ref{eq8}) with the probability $P_l(m|\k)$, we see
that Eq. (\ref{eq8}) is the first item in an unclosed hierarchy of equations,
whose generic element reads:
\beq
P_n(m|\k)=P_1(m|\k)+\sum_{l=1}^{n-1}\hat P_l(l|\k)[P_l(m|\k)-P_l(m|\k l)],
\label{eq9}
\eeq
where $\hat P_l(l|\k)=P_l(l|\k)[1-P_l(l|\k)]^{-1}$.

A natural strategy could be, at this point, perturbation theory around 
the ''ground states''
$P^\smalze_n(m)=P_1(m)$ and $P^\smalze_n(m|\k)=P_1(m|\k)$. 
Notice that substituting $P_k(k)\to P^\smalze_k(k)$ in Eq. (\ref{eq5}) and
sending $n\to\infty$ would lead to exponential scaling of the return time 
distribution:
$$
P(R_q>n)\simeq [1-P_1(\infty)]^n\simeq\exp(-n/\bar R_q),
$$
where $\bar R_q\simeq 1/P(y>q)$ [compare with Eq. (\ref{eq3.1})] 
and we have exploited $P_1(\infty)=P(y>q)>0$.

The lowest order expressions $P^\smalze_n(m)=P_1(m)$ and $P^\smalze_n(m|\k)=P_1(m|\k)$
can be calculated explicitly from PDF's in the form:
\beq
\rho_1(n)=\frac{1}{(2\pi)^{1/2}\sigma_1(n)}
\exp\Big(-\frac{(y_n-\mu_1(n))^2}{2\sigma_1^2(n)}\Big)
\label{eq10}
\eeq
and similar expression for $\rho_1(n|\k)$. 
For large $q$, in fact, the conditions $y_0,y_{k_i}>q$ in $P_1(\k)$ and $P_1(n|\k)$
can be replaced by $y_0,y_{k_i}=q$ and this guarantees Gaussian statistics.
The conditional mean and variance
in Eq. (\ref{eq10}) can be written in the following form (see e.g. \cite{olla04},
Appendix C):
\beq
\begin{array}{ll}
\mu_1(n|\k)=q\sum_{ij}C_iD_{ij},
\\
\sigma_1(n|\k)=\sum_{ij}C_iD_{ij}C_j.
\end{array}
\label{eq11}
\eeq
where $C_i=C(n-k_i)$, 
$D_{ij}$ is the inverse of the matrix $C(k_i-k_j)$ and
we have defined $k_0=0$ so that now $i,j=0,1,...p$.

Unfortunately, we are going to see that the first order correction 
$P^\smalun_n(m)=\sum_{l=1}^{n-1}\hat P_1(l)[P_1(m)-P_1(m|l)]$ 
diverges for $n\to\infty$, and the same occurs with $P^\smalun_n(m|\k)$.
As expected, exponential scaling does not appear to be an appropriate guess for
the asymptotic behavior of the return time distribution.

Let us prove this. Rearrange indices so that
$P_1(l)[P_1(m)-P_1(m|l)]\to \hat P_1(m-l)[P_1(m)-P_1(m|m-l)]$ and send
$n,m\to\infty$ with $m-n$ finite, 
so that memory of the initial condition at $t=0$ is lost.
We have in this limit:
$$
P^\smalun_n(m)=
\sum_{l=m-n+1}^\infty 
\hat P_1(\infty)[P_1(\infty)-P_1(l)].
$$
The PDF $\rho_1(\infty)$ associated with $P_1(\infty)$ is just the equilibrium PDF for $y$,
corresponding to setting in Eq. (\ref{eq10}) $\mu=0$ and $\sigma^2=1$. From Eqs. (\ref{eq11})
and (\ref{eq1}) we find:
\beq
\mu_1(l)=qAl^{-1+\epsilon},
\quad{\rm and}\quad
\sigma_1^2(l)=1-A^2l^{-2+2\epsilon},
\label{eq12}
\eeq
and, for large $l$, we can Taylor expand Eq. (\ref{eq10}):
$$
\rho_1(l)=\Big(\mu_1(l)y_l+\frac{1}{2}(y_l^2-1)(\sigma_1^2(l)-1)\Big)\rho_1(\infty).
$$
Substituting into $P_1(l)=\int_q^\infty\rho_1(l)\d y_l$ and calculating the integral
by the Laplace method, we find:
\beq
P_1(\infty)-P_1(l)=-(q^2/2)P_1(\infty) Al^{-1+\epsilon}+O(l^{-2-2\epsilon})
\label{eq13}
\eeq
where $P_1(\infty)\simeq (2\pi)^{-1/2}q^{-1}\exp(-q^2/2)$; this scaling can be compared
with the top curve in Fig. \ref{extrfig1}. Substituting into 
$\sum P_1(\infty)[P_1(\infty)-P_1(l)]$, we find that the sum is divergent 
for $\epsilon\ge 0$. Notice that the leading contribution to the scaling in Eq. 
(\ref{eq13}), the one that causes divergence of $P^\smalun_n(m)$, is the slow 
decay of the conditional mean $\mu_1(l)$ in Eq. (\ref{eq12}). 

Let us prove that also $P^\smalun_n(m|\k)$ diverges as the initial condition is sent
to $t=-\infty$.
The summand in $P^\smalun_n(m|\k)$ [see Eq. (\ref{eq9})] reads:
$$
\begin{array}{ll}
\hat P_1(l|\k)[P_1(m|\k)-P_1(m|\k l)]
\\
=
P_1(m|\k)(\hat P_1(l|\k)/P_1(l|\k))[P_1(l|\k)-P_1(l|m\k)].
\end{array}
$$
From Eq. (\ref{eq11}), for $k_1-l\gg m-k_1$: 
$\mu_1(l|\k),\mu_1(l|m\k)\simeq qC(k_1-l)\sum_{ij}D_{ij}\sim (k_1-l)^{-1+\epsilon}$
and we find again $P_1(l|\k),P_1(l|m\k)\sim [1+O(|k_1-l|^{-1+\epsilon})]P_1(\infty)$. 
This leads to the expression:
\beq
\sum_{l=1}^{n-1}\hat P_1(l|\k)[P_1(m|\k)-P_1(m|\k l)]\sim \sum_{l=1}^n l^{-1+\epsilon},
\label{eq13.1}
\eeq
which grows like $n^\epsilon$; again, 
divergence is produced by the slow decay of the conditional mean $\mu_1(l)$.

\section{{\large{$\bf \epsilon$}}-EXPANSION}
We have seen that a ''bare'' perturbation expansion of Eqs. (\ref{eq8}-\ref{eq9}),
around the zeroth order: $P^\smalze_n(m)=P_1(m)$ and $P^\smalze_n(m|\k)=P_1(m|\k)$
leads to infinities. Some kind of renormalization is necessary, however, in order
to have a workable theory, we still need that, to lowest order, the equations 
in the hierarchy (\ref{eq9}) remain decoupled. This basically fixes the
renormalization procedure.  

We separate out of the 
probabilities in Eqs. (\ref{eq8}-\ref{eq9}) a renormalized part $P^\smalR$ 
and a remnant $P^\smalN$: $P=P^\smalR+P^\smalN$, and expand
$P^\smalR=\sum_{l=0}^\infty \Delta_lP^\smalR$, where $\Delta_0P^\smalR=P_1$.
The first order renormalization to Eq. (\ref{eq8}) reads:
$$
\Delta_1P^\smalR_n(m)=\sum_{l=1}^{n-1}[\Delta_1 P^\smalR_l(l)+P_1(l)]
[P_1(m)-P_1(m|l)]_\sec,
$$
where $[P_1(m)-P_1(m|l)]_\sec$ contains the contribution leading to 
divergence of $P^\smalun_n(m)$, and, seeking 
an analogy with quantum field theory (see also \cite{chen96}), 
$\Delta_1 P^\smalR_l(l)$ could loosely be seen as a
counterterm. Analogous expressions are obtained for $\Delta_1P^\smalR_n(m|\k)$
and Eq. (\ref{eq9}). Combining with $P_1(m)$, we get the lowest order equation 
for the renormalized probability:
\beq
P^\smalR_n(m)=P_1(m)+\sum_{l=1}^{n-1}P^\smalR_l(l)
[P_1(m)-P_1(m|l)]_\sec.
\label{eq14}
\eeq
The $p$-th order renormalization,
can be expressed in the form:
\begin{eqnarray}
\Delta_pP^\smalR_n(m)&=\sum_{l=1}^{n-1}\{\Delta_pP^\smalR_l(l)
[P_l(m)-P_l(m|l)]_\sec^\smalp1
\nonumber
\\
&+P^{\smalR,p}_l(l)\Delta_{p-1}[P_l(m)-P_l(m|l)]_\sec\},
\label{eq15}
\end{eqnarray}
and we use here a superscript to indicate the order at which each expression is 
considered: $P^{\smalR,p}=P_1+\sum_{l=1}^p\Delta_lP^\smalR$
and $[\ldots]_\sec^\smalp1=\sum_{l=0}^{p-1}\Delta_l[\ldots]_\sec$.
Equation (\ref{eq15}) can be rewritten in the more concise form,
which generalizes Eq. (\ref{eq14}): 
$$
P^{\smalR,p}_n(m)=P_1(m)+\sum_{l=1}^{n-1}P^{\smalR,p}_l(l)[P_l(m)-P_l(m|l)]_\sec^\smalp1.
$$
Again, analogous expressions hold for $\Delta_pP^\smalR_n(m|\k)$
and Eq. (\ref{eq9}). From inspection of Eqs. (\ref{eq9}) and (\ref{eq15}), we
see that, in order to renormalize $P_n(m)$ to order $p$, we have to solve the first $p$ 
renormalized equations in the hierarchy (\ref{eq9}), with the second equation solved to 
order $p-1$, the third to order $p-2$, ..., the $p$-th to first order.

Turning to the remnant, $P^\smalN$ will contain, order by order, corrections 
in the form $\hat P^\smalR_l(l)-P^\smalR_l(l)$
and $[P_1(m)-P_1(m|l)]-[P_1(m)-P_1(m|l)]_\sec$, which do not lead to divergence
in Eqs. (\ref{eq14}-\ref{eq15}) and their counterparts for (\ref{eq9}).

Pursuing the analogy with quantum field theory, we see that $\epsilon=0$ plays 
a role analogous to the upper critical dimension, which suggests us to calculate 
the renormalized probability using an $\epsilon$-expansion approach. 
We choose to keep in $[\ldots]_\sec$ only
the scaling part, and, substituting Eq. (\ref{eq13}) into (\ref{eq14}), we get:
\beq
\Delta_1P^\smalR_n(m)=-\frac{1}{2}q^2AP_1(\infty)\sum_{l=1}^{n-1}P^\smalR_l(l)(m-l)^{-1+\epsilon}.
\label{eq16}
\eeq
For $n\to\infty$, 
we approximate the sum by an integral; defining $f(z)=P^\smalR_k(k)$ with
$k=n(1-z)$, we write:
$$
\sum_{l=1}^{n-1}P^\smalR_l(l)(m-l)^{-1+\epsilon}
\simeq n^\epsilon\int_\zmin^\zmax\d z f(z)[m/n-1+z]^{-1+\epsilon},
$$
where $\zmin=1/n$, $\zmax=1-1/n$ and the factor $n^\epsilon$ comes from
$n^{-1+\epsilon}/\Delta z$ with $\Delta z=1/n$ the discrete increment in the integral.
The interesting case is $m=n$. Integrating by parts and expanding 
in $\epsilon$:
$$
n^\epsilon\int_\zmin^\zmax\d z f(z)z^{-1+\epsilon}
=\epsilon^{-1}(n^\epsilon-1)P^\smalR_n(n)
+O(\epsilon).
$$
Substituting into Eq. (\ref{eq16}) and then into (\ref{eq14}), we obtain the result:
\beq
P^\smalR_n(n)=\Big[1+\frac{Aq^2P_1(\infty)}{2\epsilon}(n^\epsilon-1)\Big]^{-1}P_1(n)
\label{eq17}
\eeq
and, sending $n\to\infty$:
\beq
P^\smalR_n(n)=
\bar Pn^{-\epsilon}+O(\epsilon^2),
\label{eq18}
\eeq
where $\bar P=2\epsilon/(Aq^2)$.
Notice again the analogy with quantum field theory, with $\bar P$
behaving like the fixed point value of a renormalized coupling constant.
Substituting into Eq. (\ref{eq5}), we get the
stretched exponential scaling for the return time PDF:
\beq
P(R_q>n)\propto\exp(-\bar P n^{1-\epsilon}).
\label{eq19}
\eeq
The expectation that the result in \cite{newell62} extends to the return time statistics 
is therefore confirmed, and we have gained knowledge of the prefactor in the exponent.
Conversely, in the range $\epsilon<0$, no divergence would have arisen in Eqs.
(\ref{eq13}-\ref{eq13.1}), so that no renormalization would have been necessary.
Hence, the bare perturbation theory would have been appropriate, with the ground
state $P^\smalze_n(n)\to P_1(\infty)$ leading to exponential scaling for the
return time distribution.

\begin{figure}
\begin{center}
\includegraphics[draft=false,width=9cm]{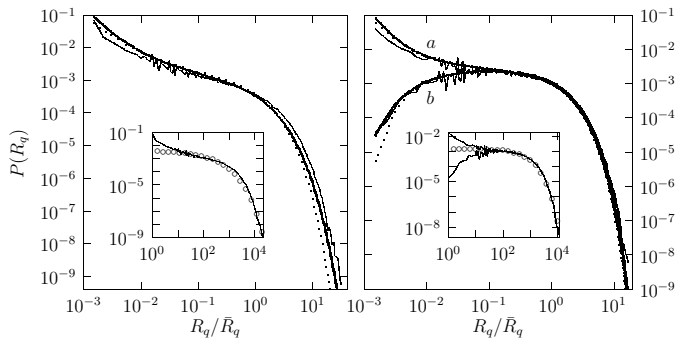}
\caption{Return time probability for $\epsilon=0.5$ (left) and 
$\epsilon=0.2$ (right, curve $a$).
Curve $b$ on the right refers to exit from the initial condition $y(0)=-q$, again 
for $\epsilon=0.2$. In all cases $\Delta=2$ and $q=3$, 
corresponding to $\bar R_q\simeq [P_1(\infty)]^{-1}\simeq 677$.
Thin lines: numerical integration of $y(t)=\sum_n x_n(t)$, using Eq. (\ref{eq3}) with
$N=24$. Dotted lines: theory using Eqs. (\ref{eq5},\ref{eq17}). Heavy lines: 
theory using Eqs. (\ref{eq27}-\ref{eq28})). 
Insert: same results without rescaling, superimposed with
the stretched exponential fits (circles): $0.004\exp(-0.1R_q^{0.5})$ and 
$0.0016\exp(-0.007 R_q^{0.8})$.
}
\label{extrfig2}
\end{center}
\end{figure}
We can use Eqs. (\ref{eq5},\ref{eq17}) to study the approach to the asymptotic 
regime (\ref{eq19}). In this transient regime, it is appropriate to use
the expression for $P_1(n)$ that is
obtained from Eqs. (\ref{eq3},\ref{eq11}), rather than 
from the asymptotic formula (\ref{eq12}).
We see in Fig. \ref{extrfig2} that the analytical approximation 
does a pretty good job even up to $\epsilon=0.5$, which is consequence of the
fact that the first correction to $\Delta_1P^\smalR_n(n)$ arises only at 
$O(\epsilon)$. 

Is this stretched exponential scaling? Actually it is possible to fit $P(R_q)$
with stretched exponentials, but the fit does not match Eq. (\ref{eq19}).
Using Eq. (\ref{eq3}), in the two cases $\epsilon=0.5$ and $\epsilon=0.2$,
we evaluate $A\simeq 0.5$ and $A\simeq 0.4$; for $q=3$, this would 
correspond to $\bar P\simeq 0.2$ in both cases, which goes from a factor of two
to two orders of magnitude away from the fits in Fig. \ref{extrfig2}.

What is happening is that Eq. (\ref{eq17}) reaches its asymptotic limit
(\ref{eq18}) only for $n\gtrsim [\bar P/P_1(\infty)]^{1/\epsilon}$, which 
grows very rapidly with $q$ and $1/\epsilon$. To have an idea, 
for $\epsilon=0.2$ and $q=3$, we would have $P_1(\infty)\simeq 0.0015$ 
and $[\bar P/P_1(\infty)]^{1/\epsilon}\sim 10^{10}$. Clearly such long return
times would occur only with vanishingly small probability.

\subsection{Higher order corrections}
Some idea of the higher orders in the perturbative expansion could be obtained
studying Eq. (\ref{eq15}) for $p=2$. We focus on the case $m=n$ and start by 
analyzing whether the second line in Eq. (\ref{eq15}), 
$\Delta_1[P_l(n)-P_l(n|l)]$ gives a secular contribution. From Eq. (\ref{eq14}),
we thus have to evaluate:
\beq
\begin{array}{ll}
\Delta_1P_l(n)=\sum_{j=1}^{l-1}\hat P^\smalR_j(j)[P_1(n)-P_1(n|j)]_\sec
\\
\Delta_1P_l(n|l)=\sum_{j=1}^{l-1}\hat P^\smalR_j(j|l)[P_1(n|l)-P_1(n|lj)]_\sec
\end{array}
\label{eq21}
\eeq
From Eq. (\ref{eq13}) and the argument leading to (\ref{eq13.1}), we see that 
$[P_1(n)-P_1(n|j)]_\sec\sim |n-j|^{-1+\epsilon}$ and 
$[P_1(n|l)-P_1(n|lj)]_\sec\sim |l-j|^{-1+\epsilon}$.
We know from Eq. (\ref{eq18}) that $P^\smalR_j(j)\sim j^{-\epsilon}$; we still
need $P^\smalR_j(j|l)$. We can repeat the steps from Eq. (\ref{eq16}) to (\ref{eq17}),
substituting $\sum_{j=1}^{l-1}P^\smalR_j(j|l)(l-1)^{-1+\epsilon}$ in the sum
of Eq. (\ref{eq16}) and the final result is again 
$$
P^\smalR_j(j|l)\sim j^{-\epsilon}.
$$
Indicating
$j=n(1-z)$ and $f(z)=P^\smalR_j(j),P^\smalR_j(l|j)$,
the leading order behavior for $l/n\to 0$ of the
sums in Eq. (\ref{eq21}) is therefore: 
$$
l^\epsilon\int_0^1f(z)[n/l-1+z]^{-1+\epsilon}
\simeq ln^{-1+\epsilon}.
$$
We see that no divergences are present at small $l$ in the sum
\beq
\sum_{l=1}^{n-1}P^\smalR_l(l)\Delta_1[P_l(n)-P_l(n|l)],
\label{eq23}
\eeq
so that $[P_l(n)-P_l(n|l)]_\sec=0$ and no renormalizations are necessary to this
order [see Eq. (\ref{eq15})].

We consider now the remnant, which receives contributions at both orders $p=1$ and
$p=2$. The only contribution surviving for $n\to\infty$ turns out to be that at
order $p=2$, produced by the sum in Eq. (\ref{eq23}). The sum is dominated by
$l\to n$; hence: $\Delta_1P_l(n)\sim P^\smalR_n(n)-P_1(n)\sim -P_1(n)$,
$\Delta_1P_{n-1}(n|n-1)\sim -P_1(n|n-1)$, and, for sufficiently large $\Delta$,
$P_1(n|n-1)\gg P_1(n)$. For $n\to\infty$, $P_1(n|n-1)=P_1(1)$ and we estimate:
\beq
P^\smalN_n(n)\sim P^\smalR_n(n)P_1(1).
\label{eq24}
\eeq
Thus, the validity of the renormalized expansion rests on the smallness
of the exit probability after one step $\lambda=P_1(1)$, that behaves
like an expansion parameter for the theory beside $\epsilon$. In order
for the theory to work, it is then necessary that 
the sampling constant $\Delta$ be sufficiently 
large. However, this appears to be a rather weak
constrain, as, already for $\Delta=1$, $q=3$ and $\epsilon=0.5$:
$\lambda\simeq 0.15$.

Substituting into Eq. (\ref{eq15}) for $p=3$, we see that
$P^\smalN_n(n)$ contributes to $\Delta_2[P_l(n)-P_l(n|l)]_\sec$ and to
renormalization of $P_n(n)$, while terms like $P_l(lji)$ contribute to 
$P^\smalN_l(n|l)$ with the same mechanism that lead to Eq. (\ref{eq24}).
This suggests that the higher order renormalizations to $P_n(n)$ are 
$O(\lambda^n)$ corrections to prefactors in Eqs. (\ref{eq17}-\ref{eq19}), 
while the exponent in (\ref{eq18}) should remain invariant. 
This exponent depends in fact only on the part of
$P_l(n|\k)-P_l(n|\k)|_{t_0\to-\infty}$ with the slowest decay,
which is $\propto |t_0|^{-1+\epsilon}$ for all $p$ [see discussion leading 
to Eq.  (\ref{eq13.1})].

\section{PDF STRUCTURE}
The key element of the analysis carried on so far is that the conditioned return probability
$P^\smalR_n(n)\simeq\bar Pn^{-\epsilon}$ [see Eq. 
(\ref{eq18})] goes to zero for $n\to\infty$. Recalling the discussion at the end 
of section \ref{section2}, it is this behavior, in contrast to that produced by 
bare perturbation theory $P_n(n)\to P_1(\infty)>0$, that leads to anomalous scaling 
of the return time distribution. This means that the PDF $\rho_n^\smalR(n)$ determining 
$P^\smalR_n(n)$ through the relation $P^\smalR_n(n)=\int_q^\infty\rho^\smalR_n(n)\d y_n$,
must have vanishing tails at $y_n>q$. The fact that for $n\to\infty$ 
$\rho^\smalR_n(n)\ne \rho(y_n)$, is the signature of the long memory of the process.

An equation for the PDF $\rho_n(n)$ could be derived repeating the steps leading to
Eq. (\ref{eq8}):
$$
\rho_n(m)=\rho_1(m)+\sum_{l=1}^{n-1}\hat P_l(l)[\rho_l(m)-\rho_l(m|l)],
$$
with obvious definitions for the various PDF's appearing in the formula. Taking moments,
we obtain equations for the conditional mean:
\beq
\mu_n(m)=\mu_1(m)+\sum_{l=1}^{n-1}\hat P_l(l)[\mu_l(m)-\mu_l(m|l)],
\label{eq25}
\eeq
and similarly for the higher moments of $\rho_n(n)$. The perturbative analysis of
Eq. (\ref{eq25}) is identical to that of Eq. (\ref{eq8}). 
Using Eqs. (\ref{eq11}-\ref{eq12}), we see that the same pattern of
divergences is produced: 
$$
[\mu_1(n)-\mu_1(n|l)]_\sec=-qA|n-l|^{-1+\epsilon}
$$
and this confirms the role of the conditional mean 
$\mu_1(n)=\langle y_n|y_0>q\rangle$ in the renormalization procedure. Generalizing
Eq. (\ref{eq25}) to the higher moments $M_{n,p}(m)=\int\rho_n(m)y_m^p\d y_m$, 
in fact, it
is possible to see that $M_{1,p}(n)-M_{1,p}(n|l)\sim |n-l|^{-p(1-\epsilon)}$ and 
for $\epsilon<0.5$, the higher moment equations do not have divergent behaviors.

\begin{figure}
\begin{center}
\includegraphics[draft=false,width=6.5cm]{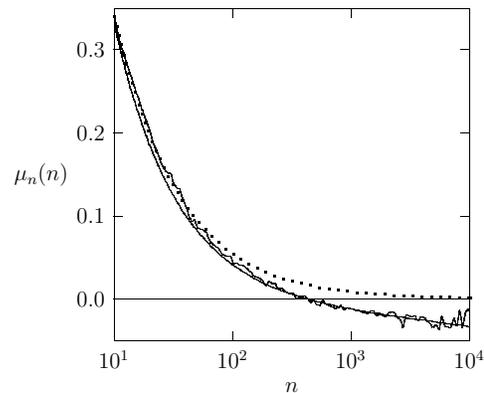}
\caption{
Scaling for the conditional mean $\mu_n(n)$ for $q=3$, $\epsilon=0.2$ and $\Delta=2$.
Dotted line: ''bare result'' $\mu_1(n)$; heavy line: renormalized result $\mu^\smalR_n(n)$
using Eq. (\ref{eq27}); thin line: numerical simulation using the same parameters of 
Fig. \ref{extrfig2}.
}
\label{extrfig3}
\end{center}
\end{figure}

Using Eq. (\ref{eq25}), it is possible to renormalize (\ref{eq24}) using the
same procedure leading from Eq. (\ref{eq16}) to (\ref{eq18}). This leads to 
the result:
\beq
\mu^\smalR_n(n)=\mu_1(n)-(q/\epsilon)(n^\epsilon-1)P^\smalR_n(n)
\to
-2/q,
\label{eq27}
\eeq
which is confirmed in Fig. \ref{extrfig3} [as in Fig. \ref{extrfig2} with $P_1(n)$, 
the expression from Eq. (\ref{eq11}) is adopted here for $\mu_1(n)$]. Notice the
constant value for $n\to\infty$ of $\mu^\smalR_n(n)$, much larger than the 
value $-qP(y>q)$ that would be obtained subtracting from the equilibrium PDF
the values of $y$ above threshold.

Clearly, knowledge of the conditional mean $\mu^\smalR_n(n)$ is not sufficient by itself
to guarantee vanishing PDF tails at $y>q$. However, using such a simple approximation
for $\rho_n(n)$ as:
\beq
\rho_n(n)\simeq (2\pi)^{-1/2}\exp\{-(y_n-\mu^\smalR_n(n))^2\},
\label{eq28}
\eeq
with $\mu^\smalR_n(n)$ given by Eq. (\ref{eq27}), and substituting into
$P_n(n)=\int_q^\infty\rho_n(n)\d y_n$ and in (\ref{eq5}), produces results in 
Fig. \ref{extrfig2} in some case better than from Eq. (\ref{eq17}).
One reason for this is the better scaling properties of $\mu_1(n)-\mu_1(n|l)\simeq
-\mu_1(n-l)$ as compared with $P_1(n)-P_1(n|l)$ (see Fig. \ref{extrfig1}).

\section{THE NON-GAUSSIAN CASE}
We have seen that the scaling of the return time distribution depends on the 
tails of the conditioned PDF $\rho_n(n)$. These in turn are determined
by the behavior of a bulk quantity like the conditional mean 
$\mu_1(n)=\langle y_n|y_0>q\rangle$. It seems therefore that it is not the extreme
statistics of the process, rather, its correlation structure that determines
the return time distribution. It is natural at this point to question whether
Gaussian statistics is strictly necessary for (quasi) stretched exponential scaling. 

We examine the conditions under which $\rho_1(\infty)-\rho_1(l)$ 
and therefore also the difference $P_1(\infty)-P_1(l)$ in Eq. (\ref{eq13}), scale like 
$\mu_1(l)\sim l^{-1+\epsilon}$. We provide a sufficient condition for this scaling in the
form of a requirement of weak correlation between scales in the process.

Let us write $y_0=x_0+z$ and $y_l=x_l+z$, with $z$ the result on $y$ of some
low-pass filtering at scale $l$ in the region in exam. 
From Eqs. (\ref{eq1}-\ref{eq2}), we have, for large $l$: 
$\langle z^2\rangle\sim l^{-1+\epsilon}$. We can obtain
$\rho_1(l)$ and $\rho_1(\infty)$ from $\rho(y)$ and $\rho(y,y';l)$, which are
respectively the equilibrium PDF for $y$ and the joint PDF that $y_l=y$ and $y_0=y'$.
Indicating by $\rho_<$ and $\rho_>$ the PDF's for the low-pass filtered signal $z$ and 
for $x=y-z$, we can write:
\beq
\begin{array}{ll}
\rho(y,y',l)&=\int\d z\rho_<(z)\rho_>(x,x';l|z),
\\
\rho(y')&=\int\d z\rho_<(z)\rho_>(x'|z).
\end{array}
\label{eq29}
\eeq
We can introduce a function $g(y,y',z,l)$ parameterizing the correlation between
the small scale components $x_l=y_l-z$ and $x_0=y_0-z$:
$$
\rho_>(x,x';l|z)=[1+g(y,y',z,l)]\rho_>(x|z)\rho_>(x'|z).
$$
If correlation between scales are weak and $l$ is large, we can
Taylor expand $\rho_>(x|z)$ and $\rho(x'|z)$ around $z=0$ and consider
$g(y,y',z,l)$ a small quantity. With these substitutions, Eq. (\ref{eq29})
becomes:
$$
\begin{array}{ll}
\rho(y,y';l)&\simeq\rho_>(y|0)\rho_>(y'|0)
[1+\langle g|y,y';l\rangle
+A\langle z^2\rangle],
\\
\rho(y')&\simeq\rho_>(y'|0)[1+B\langle z^2\rangle],
\end{array}
$$
where $A=A(y,y')$ and $B=B(y')$. From here, we obtain finally:
\beq
\rho(y|y';l)\simeq\rho(y)[1+\langle g|y,y';l\rangle+(A-B)\langle z^2\rangle].
\label{eq31}
\eeq
The two contributions 
to $\rho_1(l)-\rho_1(\infty) \simeq \rho(y|y';l)-\rho(y)$ are deeply
different in nature. The term $(A-B)\langle z^2\rangle$ is the direct additive
contribution from the long time-scale fluctuations, while $\langle g|y,y';l\rangle$
probes long time correlations of the small scale fluctuations. The last contirbution
and the direct coupling between fluctuations at different scales
\cite{benzi98}, are typically associated with a multifractal structure of the signal 
\cite{arneodo98}. 
Dominance of the additive contribution $\langle z^2\rangle\sim l^{-1+\epsilon}$, 
indicates therefore absence of multifractal properties in the signals.
The difference $P_1(\infty)-P_1(l)$ in Eq. (\ref{eq13}) will behave in this case
as if $y$ were Gaussian, and $P(R_q)$ should become a
stretched exponential in the large $R_q$ limit. 

The fact that fractal objects are not processes with a stretched exponential 
distribution of return times is not a surprise. 
This is easy to see in the case of a middle-third Cantor set, 
in which the return times can be identified with the ''holes'' in the measure: 
at the $n$-th generation 
there are $2^{n-1}$ holes of length $R_q=3^{-n}$ and this gives the power-law
distribution $P(R_q)\propto R_q^{-D}$ with $D=\ln 2/\ln 3$ the fractal dimension
of the set. Actually, there have been some recent attempts to characterize
multifractal sets by a return time spectrum, beside the more
standard singularity and dimension spectra \cite{hadyn02,roux04}.

\begin{figure}
\begin{center}
\includegraphics[draft=false,width=6.5cm]{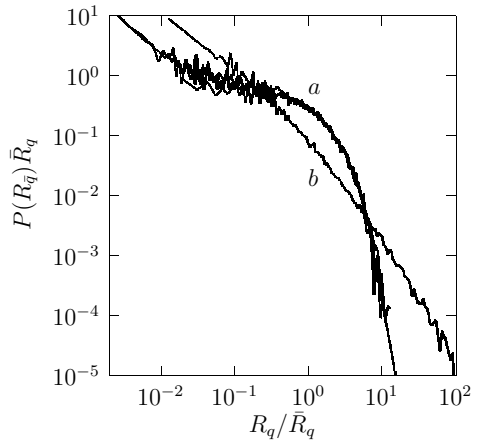}
\caption{
Return time probability for different non-Gaussian processes. In all cases,
$\epsilon=0.2$ and $\Delta=1$. Case $a$: three non-multifractal signals with
different types of non-Gaussianity; case $b$: multifractal signal.
}
\label{extrfig4}
\end{center}
\end{figure}

The prediction that the return time statistics is dominated by the correlation 
structure of the process is confirmed by numerical simulation of non-Gaussian 
power-law correlated processes, as illustrated in Fig. \ref{extrfig4}.
The four curves 
in figure are all characterized by the same power-law correlation with 
$\epsilon=0.2$, but are generated with different forms of intermittency
in the sub-processes $x_n(t)$ in $y(t)=\sum_n x_n(t)$.
The intermittency is generated letting the noise amplitude in the Langevin equation
governing each process fluctuate on the time scale of the process:
$$
\dot x_n(t)=-\tau_n x_n(t)^{-1}+b_n(t)\xi(t),
\quad
\langle\xi(t)\xi(0)\rangle=\delta(t)
$$
with $\langle b_n^2\rangle\simeq 2\sigma_{x_n}^2/\tau_n$. In the multifractal case,
the noise amplitude fluctuations are generated by a multiplicative process
of the kind utilized in \cite{arneodo98}. In the other cases, the noise amplitude
fluctuations are independent. In the first case, the fluctuations are tuned to produce
the same kurtosis $\langle y^4\rangle\simeq 12.5$ as in the fractal case. In the second
case, the intermittency grows with scale [which would produce non-trivial scaling of 
the higher diffusion exponents for $r(t)=\int_0^ty(\tau)\d\tau$]. In the third case,
the signal is Gaussian. As expected, the return
time distributions of the non-multifractal signals collapse on one another if
one rescales with $\bar R_q$, while the multifractal one leads
to a distribution that is closer to a power law.

\section{PERMANENCE TIME DISTRIBUTION}
Let us conclude the analysis turning to the permanence times and verifying that 
their distribution is characterized by stretched exponential scaling, as 
extension of the theory in \cite{newell62} would suggest. The analysis is limited
to the Gaussian case. The analog of Eq. (\ref{eq4}) in the case of the permanence 
times $S_q$ 
reads:
$$
P(S_q>t) =P^{-1}(y>q)P(y(\tau)>q,\tau\in [0,t]),
$$
where the factor $P^{-1}(y>q)$ gives the condition that the 
stochastic variable is initially above threshold. Let us isolate in
$y(\tau)$ its average in $[0,t]$:
$y(\tau)=z+x(\tau)$, where $z=t^{-1}\int_0^ty(\tau)\d\tau$.
We then obtain:
\beq
\begin{array}{ll}
&P(S_q>t) =P^{-1}(y>q)
\\
\times&\int_q^\infty\d z\rho_<(z)
P(x(\tau)>q-z,\tau\in [0,t]|z).
\end{array}
\label{eq35}
\eeq
The PDF for $z$, for large $t$, is obtained eliminating frequencies 
$|\omega|>t^{-1}$ in the power spectrum for $y$;  from Eq. (\ref{eq2}): 
$\langle z^2\rangle\sim\int_{|\omega t|<1} C_\omega\d\omega\sim t^{-1+\epsilon}$.
In the same limit, 
the condition on $z$ in 
$P(x(\tau)>q-z,\tau\in [0,t]|z)$ can be disregarded
for Gaussian statistics. 

To evaluate Eq. (\ref{eq35}), we consider the simpler problem of discrete sampling in time:
$t\to t_n=n\Delta$. This allows us to write:
$$
1>P(x(\tau)>q-z,\tau\in [0,t])>[P(x>q-z)]^{t/\Delta}.
$$
Substituting into Eq. (\ref{eq35}), the first inequality allows us to write,
to leading order in $q$ and $t$:
\beq
P(S_q>t)<\exp(-Kq^2t^{1-\epsilon}),
\label{eq36}
\eeq
where we have estimated $\rho_<(z)\sim\exp(-Kz^2t^{1-\epsilon})$
and then, for large $q$: $P(z>q)\sim\exp(-Kq^2t^{1-\epsilon})$.

\begin{figure}
\begin{center}
\includegraphics[draft=false,width=6.5cm]{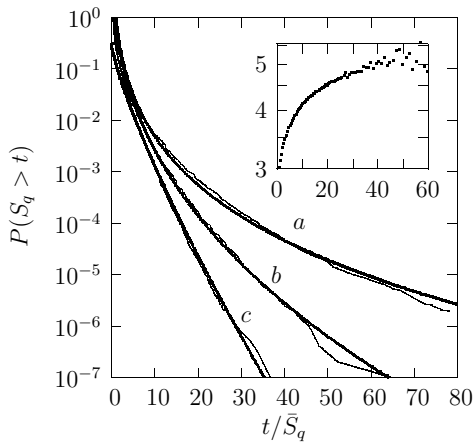}
\caption{
Permanence time distribution from numerical simulation of a power-law correlated time series 
(thin lines)
and stretched exponential fit (heavy lines) for $q=3$ and three different values of $\epsilon$:
$(a)$ $\epsilon=0.8$; $(b)$ $\epsilon=0.4$; $(c)$ $\epsilon=0.2$. Insert: scaling of
$\langle y_{\scriptscriptstyle{\rm MAX}}|S_q\rangle$ vs. $S_q$ for $\epsilon=0.4$.
}
\label{extrfig5}
\end{center}
\end{figure}

Passing to the second inequality, making the substitution in Eq. (\ref{eq35}):
$P(x(\tau)>q-z,\tau\in [0,t])\to [P(x>q-z)]^{t/\Delta}$,
the integrand in that formula will take the form:
$$
\exp\{-Kz^2t^{1-\epsilon}+(t/\Delta)\ln[1-P(x<q-z)]\}.
$$
For $t/\Delta\to\infty$, the contribution to the integral
will be from values of $z$ for which $\ln[1-P(x<q-z)$
is small, i.e. $z-q$ is large; we can approximate:
$$
\ln[1-P(x<q-z)]\sim 
-\exp(-K(q-z)^2t^{1-\epsilon}).
$$
The integrand in Eq. (\ref{eq35}) will then take the form
$$
\exp\{-Kz^2t^{1-\epsilon}-K't\exp(-K(q-z)^2t^{1-\epsilon})\},
$$
which is peaked, for $t\to\infty$, at $z\simeq q+\sqrt{\epsilon\ln t}$. Estimating the
integral in Eq. (\ref{eq35}) by steepest descent gives the result:
$\exp(-K\epsilon t^{1-\epsilon}\ln t)$. Combining with Eq. (\ref{eq36}), we
obtain the bound, valid to leading order in $q$ and $t$:
\beq
\exp(-K\epsilon t^{1-\epsilon}\ln t)<P(S_q>t)<\exp(-Kq^2t^{1-\epsilon}),
\label{eq37}
\eeq
which is similar in form to the one in \cite{newell62}. 
Contrary to the case of the return times, the value of the sampling constant $\Delta$ is
not crucial to the theory. This is confirmed by the fact that, in the present limit,
all dependence on $\Delta$, accounted for by $K'$, disappears.
We compare in Fig. \ref{extrfig5} with the result of numerical simulation using Eq. (\ref{eq3})
and see that stretched exponential scaling is compatible with the permanence time 
distributions in the range considered.

Notice that replacing the upper bound in Eq. (\ref{eq37}) with equality would imply:
$$
P(S_q>t)\sim P(z>q)\sim P(y>qt^{1-\epsilon}),
$$
in other words, the probability of a single peak of height $qS_q^{1-\epsilon}$ would be
the same as that of a permanence $S_q$. This is not surprising: given an 
initial condition $y(0)\gg q\gg 1$, from Eqs. (\ref{eq10},\ref{eq12}),
$\rho(y(t)|y(0))$ would be narrowly peaked around $y(0)\tilde A t^{-1+\epsilon}$
and the time it takes to $y(t)$ to go below $q$ would be 
$t\sim (\tilde A y(0)/q)^\frac{1}{1-\epsilon}$. Substituting $y(0)=qS_q^{1-\epsilon}$,
we obtain precisely $t\sim S_q$. As confirmed also in Fig. \ref{extrfig5}, one expects 
therefore that longer permanences above threshold be associated with higher peaks.

\section{CONCLUSION}
The analysis carried on this paper confirms the observation in 
\cite{bunde03,altmann05,eichner06} (among others) that long
correlations in stochastic processes lead to return time 
distributions with scaling close to stretched exponential.
Similar properties, albeit with different mechanisms, are confirmed for the
permanence time distribution [see Eq. (\ref{eq37}).

These results are consistent with the extension 
to thresholds different from zero, of the bounds derived
in \cite{newell62} on no zero-crossing probabilities.
The same analysis also suggests that
what can be observed in experimental time series is
only a very slow transient regime. The stretched exponential 
scaling predicted in \cite{newell62} is achieved only as an
asymptotic limit, requiring exceedingly long return times and
vanishing probabilities. However, as the theory is based on an 
$\epsilon$-expansion, it is not ruled out that the asymptotic
limit may occur earlier in the range $\epsilon\to 1$, where
new physics may become important.

From the practical point of view, it is probably irrelevant whether
a return time distribution that can be fitted by a stretched exponential
is really a stretched exponential. 
More important, 
the present theory provides approximate expressions for the return time distribution,
valid for any value of the argument and working well up to $\epsilon\simeq 0.5$, i.e. the
middle of the range considered
[see Eqs. (\ref{eq5}) and (\ref{eq18}) or 
(\ref{eq27}-\ref{eq28})]. 
The theory is limited to discrete
sampling and is actually in the form of a double expansion 
in $\epsilon$ and the parameter $\lambda=P(y_1>q|y_0>q)$,
which is heavily dependent on the sampling constant $\Delta$
[see Eq. (\ref{eq24})]. However, for large enough $q$, in the range of $\epsilon$
in which the theory works, considering $\Delta$ in the 
scaling range for the correlation $C(t)$ appears to be sufficient.

Another important fact is that, although 
the approximate expressions in the present paper are derived in the 
Gaussian case, they continue to be valid for non-Gaussian
processes, provided one rescales the relevant quantities by the mean
return time $\bar R_q$: $R_q\to R_q/\bar R_q$ and
$P(R_q)\to P(R_q)\bar R_q$. 
Basically, only multifractal processes are excluded.
(The importance in this context of rescaling by $\bar R_q$ was first
pointed out in \cite{bunde03}). 

The mechanism for the non-exponential scaling of the
return time distribution appears to be that trajectories 
originating from an above-threshold event are distributed around a mean
$\langle y(t)|y(0)>q\rangle$ that decays very slowly with time. 
When imposing that the trajectories remain below threshold 
up to the return time $R_q$, loosely speaking, this slow decay 
produces correlations between the conditions 
of no-exit at different times that cannot be treated as independent. 
The important point is that, as long as no slower scaling quantities are
characterizing the process (due e.g. to multifractality), this 
mechanism will be insensitive to whether or not the process is Gaussian
[see Eq. (\ref{eq31}) and following discussion]. This insensitivity on
the tail structure of the statistics was recently observed in 
\cite{eichner07}.

From the conceptual point of view, it is interesting that the cumulative
effect of the correlation between the below-threshold conditions becomes 
manifest through secular behaviors in the equations for the evolution of 
the trajectory distributions. 
It is also interesting that the most natural way to treat
these behaviors is renormalization, with a strategy 
similar to \cite{chen96}. In particular, the stretched exponential 
scaling limit appears to be associated with the fixed point of
the renormalized theory. The range $\epsilon<0$, corresponding
to a process with finite correlation time 
and exponential scaling for the return distribution,
conversely, is associated with a trivial theory that does not require 
renormalization.

It is to be noticed that equations similar to the ones considered in the present
paper arise in the context of avalanche models \cite{marsili98}, where
they have been treated as well within an $\epsilon$-expansion approach.

\acknowledgements
I whish to thank Cecilia Pennetta for interesting and helpful conversation.

\end{document}